\documentclass{ws-mpla}
\usepackage[super]{cite}
\usepackage{graphicx}
\usepackage{color}
\usepackage{CJKutf8}
\begin{document}
\def\Journal#1#2#3#4{{#1} {\bf #2}, #3 (#4)}
\def\AHEP{Advances in High Energy Physics.} 
\def\ARNPS{Annu. Rev. Nucl. Part. Sci.} 
\def\AandA{Astron. Astrophys.} 
\def\ANP{Ann. Phys.}
\def\APJ{Astrophys. J.}
\def\APJS{Astrophys. J. Suppl}
\def\COMR{Comptes Rendues}
\def\CQG{Class. Quantum Grav.}
\def\CPC{Chin. Phys. C}
\def\EPJC{Eur. Phys. J. C}
\def\EPL{EPL}
\def\FSPAS{Front. Astron. Space Sci.}
\def\IJMPA{Int. J. Mod. Phys. A}
\def\IJMPE{Int. J. Mod. Phys. E}
\def\IJTP{Int. J. Theor. Phys.}
\def\JCAP{J. Cosmol. Astropart. Phys.}
\def\JHEP{J. High Energy Phys.}
\def\JETPL{JETP. Lett.}
\def\JETPUSSR{JETP (USSR)}
\def\JPG{J. Phys. G} 
\def\JPCS{J. Phys. Conf. Ser.} 
\def\JPGNP{J. Phys. G: Nucl. Part. Phys.} 
\def\MPLA{Mod. Phys. Lett. A}
\def\NIMA{Nucl. Instrum. Meth. A.}
\def\NATU{Nature}
\def\NCA{Nuovo Cimento}
\def\NJP{New. J. Phys.}
\def\NPB{Nucl. Phys. B}
\def\NPBOLD{Nucl. Phys.}
\def\NPBSUPPL{Nucl. Phys. B. Proc. Suppl.}
\def\PL{Phys. Lett.}
\def\PLB{{Phys. Lett.} B}
\def\PMCA{PMC Phys. A}
\def\PREP{Phys. Rep.}
\def\PPNP{Prog. Part. Nucl. Phys.}
\def\PLBOLD{Phys. Lett.}
\def\PAN{Phys. Atom. Nucl.}
\def\PRL{Phys. Rev. Lett.}
\def\PRD{Phys. Rev. D}
\def\PRC{Phys. Rev. C}
\def\PR{Phys. Rev.}
\def\PTP{Prog. Theor. Phys.}
\def\PTEP{Prog. Theor. Exp. Phys.}
\def\PTPS{Prog. Theor. Phys. Suppl}
\def\RMP{Rev. Mod. Phys.}
\def\RPP{Rep. Prog. Phys.}
\def\SJNP{Sov. J. Nucl. Phys.}
\def\SCIENCE{Science}
\def\TNYAS{Trans. New York Acad. Sci.}
\def\ZETP{Zh. Eksp. Teor. Piz.}
\def\ZFPH{Z. fur Physik}
\def\ZPC{Z. Phys. C}

\markboth{Eito Nagao, Yuta Hyodo and Teruyuki Kitabayashi}{Scaling in the minimal extended seesaw model}

\catchline{}{}{}{}{}

\title{Scaling in the minimal extended seesaw model}

\author{Eito Nagao}

\address{Graduate School of Science, Tokai University, 4-1-1 Kitakaname, Hiratsuka, Kanagawa 259-1292, Japan}

\author{Yuta Hyodo and Teruyuki Kitabayashi}

\address{Department of Physics, Tokai University, 4-1-1 Kitakaname,\\ Hiratsuka, Kanagawa 259-1292, Japan \\ Corresponding author: teruyuki@tokai.ac.jp}

\maketitle

\pub{Received (Day Month Year)}{Revised (Day Month Year)}

\begin{abstract}
The scaling relations in the two-zero textures of the neutrino mass matrices in the minimal extended seesaw mechanism were discovered by Kumar and Patgiri. We demonstrate that some of these scaling relations can be satisfied without requiring two zero elements in the texture.

\end{abstract}

\ccode{PACS Nos.: 14.60.Pq}

\section{Introduction\label{section:introduction}}	
Since the tiny masses of neutrinos cannot be explained by the standard model of particle physics, they provide important clues for the development of theories beyond the standard model. As a result, many possible textures of the neutrino mass matrix have been proposed \cite{Frizsch2020PPNP ,Altarelli2010RMP
,Ishimori2010PTPS
,Xing2020PREP
,Xing2023RPP
}.

An important factor in determining the correct texture of the neutrino mass matrix is the so-called scaling concept in the neutrino mass matrix \cite{Mohapatra2007PLB}. In certain textures of the neutrino mass matrix, the elements of the matrix obey a scaling law, where the ratios of specific elements of the matrix are equal \cite{Mohapatra2007PLB,Blum2007PRD,Joshipura2009PLB,Yasur2012PRD,Samanta2016EPJC,Sinha2017JHEP}. The scaling relations in the two-zero textures of $4\times 4$ neutrino mass matrices within the minimal extended seesaw (MES) mechanism \cite{Barry2011JHEP} have been explored by Kumar and Patgiri \cite{Kumar2020NPB} \footnote{The phenomenology of the two-zero textures in the MES mechanism has been discussed \cite{Nath2017JHEP,Patgiri2017IJMPA,Sarma2019EPJC,Das2019NPB,Patgiri2019IJMPA,Kumar2020NPB,Das2022NPB}.}. For example, they found that in the following two-zero texture of the neutrino mass matrix
\begin{align}
M_\nu
=
\begin{pmatrix}
m_{ee}&m_{e\mu}&m_{e\tau}&m_{es} \\
m_{e\mu}&m_{\mu\mu}&m_{\mu\tau}&m_{\mu s} \\
m_{e\tau}&m_{\mu\tau}&m_{\tau\tau}&m_{\tau s} \\
m_{es}&m_{\mu s}&m_{\tau s}&m_{s s} \\
\end{pmatrix}
=
    \begin{pmatrix}
        0 &0 &*  & * \\
        0& * &* & * \\
        *& * &* & *\\
        *& * &*&* \\
    \end{pmatrix},
\end{align}
where $*$ denotes a nonzero element and a scaling
\begin{align}\label{A1corelation1}
\frac{m_{e \tau}}{m_{es}} =\frac{m_{\mu \tau}}{m_{\mu s}}= \frac{m_{\tau \tau}}{m_{\tau s}}=\frac{m_{\tau s}}{m_{ss}}=\sqrt{\frac{m_{\tau \tau}}{m_{ss}}},
\end{align}
is satisfied with the specific condition in the MES mechanism. 

We would like to highlight that the two zero elements were absent in the scaling. For instance, $m_{ee} (=0)$ and $m_{e\mu} (=0)$ are missing in Eq. (\ref{A1corelation1}). These observations suggest that some scaling relations may exist in two-zero textures even if the neutrino mass matrix does not contain two zero elements. In this study, we demonstrate that certain scaling relations identified by Kumar and Patgiri in two-zero textures can indeed be satisfied without the need for two zero elements in the texture. 

The remainder of the paper is organized as follows. Section \ref{section_scaling_two_zero} presents a brief review of the scaling in the two-zero textures in the MES mechanism \cite{Kumar2020NPB}. In Section \ref{section_general_scaling}, we demonstrate that some scaling relations do not require two zero elements in the mass matrix. Finally, Section \ref{section:summary} provides a summary of the paper.

\section{Scaling in two-zero textures \label{section_scaling_two_zero}}
The MES mechanism is an extension of the type-I seesaw mechanism with an additional gauge singlet field \cite{Barry2011JHEP}. The $4 \times 4$ Majorana neutrino mass matrix in the MES mechanism is obtained as
\begin{align}\label{massmatrix4}
M_\nu=
\begin{pmatrix}
M_DM^{-1}_{R}M^{T}_{D} & M_DM^{-1}_{R}M^{T}_{S} \\
 M^{T}_{S}(M^{-1}_{R})^{T}M^{T}_{D} &M_SM^{-1}_{R}M^{T}_{S} \\
\end{pmatrix}, 
\end{align}
where $M_D$ represents the $3\times 3$ Dirac neutrino mass matrix, $M_R$ represents the $3\times 3$ right-handed Majorana neutrino mass matrix, and $1\times 3$ matrix $M_S$ denotes the coupling of the three right-handed neutrinos with the singlet (sterile) field $S$.

Kumar and Patgiri found the scaling relations in the following nine types of two-zero textures of $M_\nu$ \cite{Kumar2020NPB}:
\begin{align}
   &A_1:
    \begin{pmatrix}
        0 &0 &*  & * \\
        0& * &* & * \\
        *& * &* & *\\
        *& * &*&* \\
    \end{pmatrix},
    \quad
    A_2:
    \begin{pmatrix}
        0& * &0  & * \\
        *& * &* & * \\
        0& * &* & *\\
        *& * &*&* \\
    \end{pmatrix},
    \quad
    B_3:
    \begin{pmatrix}
        * &0 &*  & * \\
        0& 0 &* & * \\
        *& * &* & *\\
        *& * &*&* \\
    \end{pmatrix},
    \quad
    B_4:
    \begin{pmatrix}
        * &* &0  & * \\
        *& * &* & * \\
        0& * &0 & *\\
        *& * &*&* \\
    \end{pmatrix},
    \nonumber \\ 
    &D_1:
    \begin{pmatrix}
        * &* &*  & * \\
        *& 0 &0 & * \\
        *& 0 &* & *\\
        *& * &*&* \\
    \end{pmatrix},
    \quad
    D_2:
    \begin{pmatrix}
        * &* &*  & * \\
        *& * &0 & * \\
        *& 0 &0 & *\\
        *& * &*&* \\
    \end{pmatrix},
    \quad
    F_1:
    \begin{pmatrix}
        * &0 &0  & * \\
        0& * &* & * \\
        0& * &* & *\\
        *& * &*&* \\
    \end{pmatrix},
    \quad
    F_2:
    \begin{pmatrix}
        * &0 &*  & * \\
        0& * &0 & * \\
        *& 0 &* & *\\
        *& * &*&* \\
    \end{pmatrix},
    \nonumber \\
    &F_3:
    \begin{pmatrix}
        * &* &0  & * \\
        *& * &0 & * \\
        0& 0 &* & *\\
        *& * &*&* \\
    \end{pmatrix}.
\label{Eq:textureTwoZero}
\end{align}
To realize these two-zero textures, according to Ref. \cite{Kumar2020NPB}, the candidates of the explicit form of $M_D$, $M_R$ and $M_S$ are
\begin{align}
&M^{(1)}_D =
\begin{pmatrix}
a&0&0 \\
0&0&f\\
0&h&\ell \\
\end{pmatrix},
\quad
&M^{(2)}_D =
\begin{pmatrix}
0&b&0 \\
d&0&0 \\
g&0&\ell \\
\end{pmatrix},
\quad
&M^{(3)}_D =
\begin{pmatrix}
0&b&0 \\
d&e&0 \\
0&0&\ell \\
\end{pmatrix},
\quad
&M^{(4)}_D =
\begin{pmatrix}
0&b&0 \\
d&0&0 \\
0&h&\ell \\
\end{pmatrix},
\nonumber \\
&M^{(5)}_D =
\begin{pmatrix}
a&0&0 \\
d&0&f \\
0&h&0 \\
\end{pmatrix},
\quad
&M^{(6)}_D =
\begin{pmatrix}
a&0&0 \\
0&e&f \\
0&0&\ell \\
\end{pmatrix},
\quad
&M^{(7)}_D =
\begin{pmatrix}
a&0&0 \\
0&e&0 \\
g&0&\ell \\
\end{pmatrix},
\quad
&M^{(8)}_D =
\begin{pmatrix}
0&b&0 \\
d&0&f \\
g&0&0 \\
\end{pmatrix},
\nonumber \\
&M^{(9)}_D =
\begin{pmatrix}
0&b&0 \\
0&e&f \\
g&0&0 \\
\end{pmatrix},
\quad
&M^{(10)}_D =
\begin{pmatrix}
0&b&0 \\
0&0&f \\
g&h&0 \\
\end{pmatrix},
\quad
&M^{(11)}_D =
\begin{pmatrix}
a&b&0 \\
0&e&0 \\
0&0&\ell \\
\end{pmatrix},
\quad
&M^{(12)}_D =
\begin{pmatrix}
a&0&c \\
d&0&0 \\
0&h&0 \\
\end{pmatrix},
\nonumber \\
&M^{(13)}_D =
\begin{pmatrix}
a&0&0 \\
0&e&0 \\
0&h&\ell \\
\end{pmatrix},
\quad
&M^{(14)}_D =
\begin{pmatrix}
0&0&c \\
d&0&0 \\
0&h&\ell \\
\end{pmatrix},
\quad
&M^{(15)}_D =
\begin{pmatrix}
a&b&0 \\
0&0&f \\
0&h&0 \\
\end{pmatrix},
\quad
&M^{(16)}_D =
\begin{pmatrix}
a&0&c \\
0&e&0 \\
g&0&0 \\
\end{pmatrix},
\nonumber \\
&M^{(17)}_D =
\begin{pmatrix}
a&0&0 \\
0&e&f \\
0&h&0 \\
\end{pmatrix},
\quad
&M^{(18)}_D =
\begin{pmatrix}
0&0&c \\
0&e&f \\
g&0&0 \\
\end{pmatrix},
\quad
&M^{(19)}_D =
\begin{pmatrix}
0&b&c \\
d&0&0 \\
0&0&\ell \\
\end{pmatrix},
\quad
&M^{(20)}_D =
\begin{pmatrix}
0&b&c \\
0&e&0 \\
g&0&0 \\
\end{pmatrix},
\nonumber \\
&M^{(21)}_D =
\begin{pmatrix}
0&0&c \\
0&e&0 \\
g&h&0 \\
\end{pmatrix},
\quad
&M^{(22)}_D =
\begin{pmatrix}
0&b&c \\
d&0&0 \\
0&h&0 \\
\end{pmatrix},
\quad
&M^{(23)}_D =
\begin{pmatrix}
0&b&c \\
0&0&f \\
g&0&0 \\
\end{pmatrix},
\quad
&M^{(24)}_D =
\begin{pmatrix}
0&0&c \\
d&e&0 \\
0&h&0 \\
\end{pmatrix},
\nonumber \\
&M^{(25)}_D =
\begin{pmatrix}
a&b&0 \\
d&0&0 \\
0&0&\ell \\
\end{pmatrix},
\quad
&M^{(26)}_D =
\begin{pmatrix}
a&0&0 \\
d&e&0 \\
0&0&\ell \\
\end{pmatrix},
\end{align}
\begin{align}
M^{(1)}_R =
\begin{pmatrix}
A&0&0 \\
0&D&0\\
0&0&F\\
\end{pmatrix},
\quad
M^{(2)}_R =
\begin{pmatrix}
A&B&0 \\
B&0&E\\
0&E&0\\
\end{pmatrix},
\quad
M^{(3)}_R =
\begin{pmatrix}
A&B&0 \\
B&0&0\\
0&0&F\\
\end{pmatrix},&
\end{align}
and
\begin{align}
M^{(1)}_S =
\begin{pmatrix}
0&s_2&s_3 \\
\end{pmatrix},
\quad
M^{(2)}_S =
\begin{pmatrix}
s_1&0&s_3 \\
\end{pmatrix},
\quad
M^{(3)}_S =
\begin{pmatrix}
s_1&s_2&0 \\
\end{pmatrix},
\nonumber \\
M^{(4)}_S =
\begin{pmatrix}
s_1&0&0 \\
\end{pmatrix},
\quad
M^{(5)}_S =
\begin{pmatrix}
0&s_2&0 \\
\end{pmatrix},
\quad
M^{(6)}_S =
\begin{pmatrix}
0&0&s_3 \\
\end{pmatrix},
\end{align}
respectively.

The scaling relations depend on the explicit form of $M_D$, $M_R$, and $M_S$. For example, the combination
\begin{align}
\left(M_D, M_R, M_S\right)=\left(M^{(3)}_D, M^{(2)}_R, M^{(6)}_S\right),
\end{align}
leads to an $A_1$ texture as
\begin{align}\label{A1(3)}
M_{\nu}=
\begin{pmatrix}
0 & 0 & \frac{b \ell}{F} & \frac{b s_{3}}{F}\\0 & \frac{d^{2}}{A} & \frac{\ell \left(A e - B d\right)}{A F} & \frac{s_{3} \left(A e - B d\right)}{A F}\\
\frac{b \ell}{F} & \frac{\ell \left(A e - B d\right)}{A F} & \frac{B^{2} \ell^{2}}{A F^{2}} & \frac{B^{2} \ell s_{3}}{A F^{2}}\\
\frac{b s_{3}}{F} & \frac{s_{3} \left(A e - B d\right)}{A F} & \frac{B^{2} \ell s_{3}}{A F^{2}} & \frac{B^{2} s_{3}^{2}}{A F^{2}}
\end{pmatrix},
\end{align}
and the following scaling
\begin{align}\label{correlation1}
\frac{m_{e \tau}}{m_{es}} =\frac{m_{\mu \tau}}{m_{\mu s}}= \frac{m_{\tau \tau}}{m_{\tau s}}=\frac{m_{\tau s}}{m_{ss}}=\sqrt{\frac{m_{\tau \tau}}{m_{ss}}}.
\end{align}
On the other hand, although other combinations
\begin{align}
\left(M_D, M_R, M_S\right)=\left(M^{(4)}_D, M^{(2)}_R, M^{(6)}_S\right),
\end{align}
also leads to an $A_1$ texture as
\begin{align}
M_\nu=
    \begin{pmatrix}
        0 & 0 & \frac{b \ell}{F} & \frac{b s_{3}}{F}\\
        0 & \frac{d^{2}}{A}& - \frac{B d \ell}{A F} & - \frac{B d s_{3}}{A F} \\
        \frac{b \ell}{F} & - \frac{B d \ell}{A F}  &\frac{\ell \left(2 A F h + B^{2} \ell \right)}{A F^{2}} & \frac{s_{3} \left(A F h + B^{2} \ell \right)}{A F^{2}}\\
        \frac{b s_{3}}{F} & - \frac{B d s_{3}}{A F}& \frac{s_{3} \left(A F h + B^{2} \ell \right)}{A F^{2}}&\frac{B^{2} s_{3}^{2}}{A F^{2}}  \\
    \end{pmatrix},
\end{align}
the following different scaling \textcolor{black}{relations are} obtained:
\begin{align}\label{A1correlation2}
\frac{m_{e \tau}}{m_{es}} = \frac{m_{\mu \tau}}{m_{\mu s}}, \quad \frac{m_{\mu \mu}}{m_{\mu s}} = \frac{m_{\mu s}}{m_{ss}}.
\end{align}
All the scaling relations in the two-zero textures in the MES mechanism are shown in the next section.

\section{More general scaling \label{section_general_scaling}}
A scaling that is independent of the two-zero texture can be regarded as a more general scaling than a restrictive scaling that requires two zero elements in the mass matrix. In this section, we show that some scaling relations that Kumar and Patgiri found in two-zero textures can be satisfied without requiring two zero elements in the texture. 

To achieve this, we prepare $\tilde M^{(i)}_D$ $(i=1, 2, \cdots, 26)$ that has the same texture of $M^{(i)}_D$ but has $\epsilon_D$ instead of zero. Similarly, $\tilde M^{(i)}_R$ $(i=1,2,3)$ with $\epsilon_R$ and $\tilde M^{(i)}_S$ $(i=1,2,\cdots, 6)$ with $\epsilon_S$ are prepared.\footnote{\color{black}In general, $M_D$ (or $\tilde{M}_D$), $M_R$ (or $\tilde{M}_R$), and $M_S$ (or $\tilde{M}_S$) have 9, 6 and 3 independent complex elements, respectively. Thus, the assumption in this paper, such as Eq.(\ref{Eq:ExTildeM}), is quite strong. For a more general analysis, we need to consider the generalized matrices, such as 
\begin{align}
M^{(1)}_D =
\begin{pmatrix}
a&0&0 \\
0&0&f\\
0&h&\ell \\
\end{pmatrix}
 & \rightarrow 
\tilde M^{(1)}_D =
\begin{pmatrix}
a&\epsilon_{D1}&\epsilon_{D2} \\
\epsilon_{D3}&\epsilon_{D4}&f\\
\epsilon_{D5}&h&\ell \\
\end{pmatrix},
\nonumber \\
M_R^{(1)} = 
\begin{pmatrix}
A&0 & 0\\
0 &D&0\\
0 &0 &F\\
\end{pmatrix}
& \rightarrow
\tilde{M}_R^{(1)} = 
\begin{pmatrix}
A&\epsilon_{R1} & \epsilon_{R2}\\
\epsilon_{R1} &D&\epsilon_{R3}\\
\epsilon_{R2} &\epsilon_{R3} &F\\
\end{pmatrix},
\nonumber \\
M^{(4)}_S =
\begin{pmatrix}
s_1&0&0 \\
\end{pmatrix}
&\rightarrow
\tilde M^{(4)}_S =
\begin{pmatrix}
s_1 &\epsilon_{S1}&\epsilon_{S2} \\
\end{pmatrix}, \nonumber
\end{align}
instead of Eq.(\ref{Eq:ExTildeM}). In our preliminary study, we obtained results similar to those in Table \ref{table_scaling} with such a complete generalization. We would like to intend to study this issue in more detail in the future.}
For example,
\begin{align}
M^{(1)}_D =
\begin{pmatrix}
a&0&0 \\
0&0&f\\
0&h&\ell \\
\end{pmatrix}
 & \rightarrow 
\tilde M^{(1)}_D =
\begin{pmatrix}
a&\epsilon_D&\epsilon_D \\
\epsilon_D&\epsilon_D&f\\
\epsilon_D&h&\ell \\
\end{pmatrix},
\nonumber \\
M_R^{(1)} = 
\begin{pmatrix}
A&0 & 0\\
0 &D&0\\
0 &0 &F\\
\end{pmatrix}
& \rightarrow
\tilde{M}_R^{(1)} = 
\begin{pmatrix}
A&\epsilon_R & \epsilon_R\\
\epsilon_R &D&\epsilon_R\\
\epsilon_R &\epsilon_R &F\\
\end{pmatrix},
\nonumber \\
M^{(4)}_S =
\begin{pmatrix}
s_1&0&0 \\
\end{pmatrix}
&\rightarrow
\tilde M^{(4)}_S =
\begin{pmatrix}
s_1 &\epsilon_S&\epsilon_S \\
\end{pmatrix}.
\label{Eq:ExTildeM}
\end{align}

We define $\tilde A_1$ as a type of $4\times 4$ neutrino mass matrix that becomes $A_1$ in the case of $\epsilon_D = 0$, $\epsilon_R = 0$, and $\epsilon_S = 0$. For example, the combination
\begin{align}
\left(M_D, M_R, M_S\right)=\left(M^{(3)}_D, \tilde M^{(2)}_R, M^{(6)}_S\right),
\end{align}
leads to an $\tilde A_1$ texture, since the elements 
\begin{align}\label{A1tildecorrelation1}
&m_{ee}=
\frac{\epsilon_R b^{2} \left(- A + \epsilon_R\right)}{A F^{2} - A \epsilon_R^{2} + B^{2} \epsilon_R - 2 B F \epsilon_R + \epsilon_R^{3}},
\quad
m_{e\mu}=
\frac{\epsilon_R b \left(d \left(B - F\right) - e \left(A - \epsilon_R\right)\right)}{A F^{2} - A \epsilon_R^{2} + B^{2} \epsilon_R - 2 B F \epsilon_R + \epsilon_R^{3}},
\nonumber \\
&m_{e\tau}=
\frac{b \ell \left(A F - B \epsilon_R\right)}{A F^{2} - A \epsilon_R^{2} + B^{2} \epsilon_R - 2 B F \epsilon_R + \epsilon_R^{3}},
\quad
m_{es} =
\frac{b s_{3} \left(A F - B \epsilon_R\right)}{A F^{2} - A \epsilon_R^{2} + B^{2} \epsilon_R - 2 B F \epsilon_R + \epsilon_R^{3}},
\nonumber \\
&m_{\mu \mu} = 
\frac{\epsilon_R e \left(d \left(B - F\right) - e \left(A - \epsilon_R\right)\right) + d \left(\epsilon_R e \left(B - F\right) + d \left(F^{2} - \epsilon_R^{2}\right)\right)}{A F^{2} - A \epsilon_R^{2} + B^{2} \epsilon_R - 2 B F \epsilon_R + \epsilon_R^{3}},
\nonumber \\
&m_{\mu \tau} = 
- \frac{\ell \left(d \left(B F - \epsilon_R^{2}\right) - e \left(A F - B \epsilon_R\right)\right)}{A F^{2} - A \epsilon_R^{2} + B^{2} \epsilon_R - 2 B F \epsilon_R + \epsilon_R^{3}},
\quad
m_{\mu s} =
- \frac{s_{3} \left(d \left(B F - \epsilon_R^{2}\right) - e \left(A F - B \epsilon_R\right)\right)}{A F^{2} - A \epsilon_R^{2} + B^{2} \epsilon_R - 2 B F \epsilon_R + \epsilon_R^{3}},
\nonumber \\
&m_{\tau \tau} =
\frac{\ell^{2} \left(- A \epsilon_R + B^{2}\right)}{A F^{2} - A \epsilon_R^{2} + B^{2} \epsilon_R - 2 B F \epsilon_R + \epsilon_R^{3}},
\quad
m_{\tau s} =
\frac{\ell s_{3} \left(- A \epsilon_R + B^{2}\right)}{A F^{2} - A \epsilon_R^{2} + B^{2} \epsilon_R - 2 B F \epsilon_R + \epsilon_R^{3}},
\nonumber \\
&m_{ss} = 
\frac{s_{3}^{2} \left(- A \epsilon_R + B^{2}\right)}{A F^{2} - A \epsilon_R^{2} + B^{2} \epsilon_R - 2 B F \epsilon_R + \epsilon_R^{3}},
\end{align}
are satisfied with the criteria of $A_1$ (i.e., $m_{ee}=m_{e\mu}=0$) in the case of $\epsilon_R = 0$. For example, the combination 
\begin{align}
\left(M_D, M_R, M_S\right)=\left(M^{(4)}_D, \tilde M^{(2)}_R, M^{(6)}_S\right),
\end{align}
also leads to an $\tilde A_1$ texture that has the following elements:
\begin{align}\label{A1tildecorrelation2}
&m_{ee}= \frac{ \epsilon_{R} b^{2} \left(- A +  \epsilon_{R}\right)}{A F^{2} - A  \epsilon_{R}^{2} + B^{2}  \epsilon_{R} - 2 B F  \epsilon_{R} +  \epsilon_{R}^{3}},
\quad
m_{e\mu}= \frac{ \epsilon_{R} b d \left(B - F\right)}{A F^{2} - A  \epsilon_{R}^{2} + B^{2}  \epsilon_{R} - 2 B F  \epsilon_{R} +  \epsilon_{R}^{3}} ,
\nonumber \\
&m_{e\tau}= \frac{b \left(-  \epsilon_{R} h \left(A -  \epsilon_{R}\right) + \ell \left(A F - B  \epsilon_{R}\right)\right)}{A F^{2} - A  \epsilon_{R}^{2} + B^{2}  \epsilon_{R} - 2 B F  \epsilon_{R} +  \epsilon_{R}^{3}} ,
\quad
m_{es} = \frac{b s_{3} \left(A F - B  \epsilon_{R}\right)}{A F^{2} - A  \epsilon_{R}^{2} + B^{2}  \epsilon_{R} - 2 B F  \epsilon_{R} +  \epsilon_{R}^{3}} ,
\nonumber \\
&m_{\mu \mu} =\frac{d^{2} \left(F^{2} -  \epsilon_{R}^{2}\right)}{A F^{2} - A  \epsilon_{R}^{2} + B^{2}  \epsilon_{R} - 2 B F  \epsilon_{R} +  \epsilon_{R}^{3}} , 
\quad
m_{\mu \tau} = \frac{d \left( \epsilon_{R} h \left(B - F\right) - \ell \left(B F -  \epsilon_{R}^{2}\right)\right)}{A F^{2} - A  \epsilon_{R}^{2} + B^{2}  \epsilon_{R} - 2 B F  \epsilon_{R} +  \epsilon_{R}^{3}},
\nonumber \\
&m_{\mu s} = \frac{d s_{3} \left(- B F +  \epsilon_{R}^{2}\right)}{A F^{2} - A  \epsilon_{R}^{2} + B^{2}  \epsilon_{R} - 2 B F  \epsilon_{R} +  \epsilon_{R}^{3}} ,
\nonumber \\
&m_{\tau \tau} = \frac{- h \left( \epsilon_{R} h \left(A -  \epsilon_{R}\right) - \ell \left(A F - B  \epsilon_{R}\right)\right) + \ell \left(h \left(A F - B  \epsilon_{R}\right) - \ell \left(A  \epsilon_{R} - B^{2}\right)\right)}{A F^{2} - A  \epsilon_{R}^{2} + B^{2}  \epsilon_{R} - 2 B F  \epsilon_{R} +  \epsilon_{R}^{3}} ,
\nonumber \\
&m_{\tau s} = \frac{s_{3} \left(h \left(A F - B  \epsilon_{R}\right) - \ell \left(A  \epsilon_{R} - B^{2}\right)\right)}{A F^{2} - A  \epsilon_{R}^{2} + B^{2}  \epsilon_{R} - 2 B F  \epsilon_{R} +  \epsilon_{R}^{3}} ,
\quad
m_{ss} = \frac{s_{3}^{2} \left(- A  \epsilon_{R} + B^{2}\right)}{A F^{2} - A  \epsilon_{R}^{2} + B^{2}  \epsilon_{R} - 2 B F  \epsilon_{R} +  \epsilon_{R}^{3}}.
\end{align}
We define $\tilde A_2, \tilde B_3, \tilde B_4, \tilde D_1, \tilde D_2, \tilde F_1, \tilde F_2$, and $\tilde F_3$ in the same way. 

It is worth noting that the scaling in a two-zero texture $A_1$ via $\left(M_D, M_R, M_S\right)=\left(M^{(3)}_D, M^{(2)}_R, M^{(6)}_S\right)$, see Eq. (\ref{correlation1}), is still satisfied with texture $\tilde A_1$ via $\left(M_D, M_R, M_S\right)=\left(M^{(3)}_D, \tilde M^{(2)}_R, M^{(6)}_S\right)$, see Eq. (\ref{A1tildecorrelation1}). In contrast, 
the scaling in another two-zero texture $A_1$ via $\left(M_D, M_R, M_S\right)=\left(M^{(4)}_D, M^{(2)}_R, M^{(6)}_S\right)$, see Eq. (\ref{A1correlation2}), is no longer active for the texture $\tilde A_1$ via $\left(M_D, M_R, M_S\right)=\left(M^{(4)}_D, \tilde M^{(2)}_R, M^{(6)}_S\right)$, see Eq. (\ref{A1tildecorrelation2}). Therefore, some of the scaling relations that Kumar and Patgiri found in the two-zero textures can be satisfied without requiring two zero elements in the texture.

The seven combinations of $M_D$, $M_R$, and $M_S$ can be used for making $\tilde A_1, \tilde A_2, \cdots, \tilde F_3$:
\begin{align}
\left(M_D, M_R, M_S\right) = &\left(\tilde M^{(i)}_D, M^{(j)}_R, M^{(k)}_S\right), \left(M^{(i)}_D, \tilde M^{(j)}_R, M^{(k)}_S\right), \left(M^{(i)}_D, M^{(j)}_R, \tilde M^{(k)}_S\right),  
\nonumber \\
&\left(\tilde M^{(i)}_D, \tilde M^{(j)}_R, M^{(k)}_S\right), \left(\tilde M^{(i)}_D, M^{(j)}_R, \tilde M^{(k)}_S\right), \left(M^{(i)}_D, \tilde M^{(j)}_R, \tilde M^{(k)}_S\right), 
\nonumber \\
&\left(\tilde M^{(i)}_D, \tilde M^{(j)}_R, \tilde M^{(k)}_S\right). 
\end{align}
The scaling relations in the two-zero textures $A_1, A_2, \cdots, F_3$ and the more general textures $\tilde A_1, \tilde A_2, \cdots, \tilde F_3$ are listed in Tables \ref{table_scaling} and \ref{scalingcon}. In these tables, `Type' represents the combinations of ($M_D, M_R, M_S$),  ($M_D, \tilde M_R, M_S$), and  ($M_D, M_R, \tilde M_S$) for two-zero textures, more general textures associated with $\tilde M_R$, and more general textures associated with $\tilde M_S$, respectively. It turned out that 
\begin{itemize}
\item some scaling relations in $A_1, A_2, \cdots D_2$ still survive in $\tilde A_1, \tilde A_2, \cdots \tilde D_2$ for the following two combinations:
\begin{align}
\left(M_D, M_R, M_S\right) = \left(M^{(i)}_D, \tilde M^{(j)}_R, M^{(k)}_S\right), \left(M^{(i)}_D, M^{(j)}_R, \tilde M^{(k)}_S\right),  
\end{align}
\item scaling relations in $F_1, F_2, F_3$ cannot be permitted in $\tilde F_1, \tilde F_2, \tilde F_3$.
\end{itemize}

Thus, some scaling relations that Kumar and Patgiri found in two-zero textures can be satisfied without requiring two zero elements in the texture. This is the main finding of this paper.

\begin{table}[t]
\tbl{Scaling relations in the MES.}
{\begin{tabular}{|c|c|c|c|c|}
\hline
Type& Scaling & $A_1$ & $\tilde{A}_1(\tilde M_R)$ & $\tilde{A}_1(\tilde M_S)$ \\
\hline
(1,3,1)
&
$
\frac{(m_{\tau s}m_{\mu \mu} - m_{\mu \tau}m_{\mu s} )}{m_{e\tau}}  = \frac{(m_{s s}m_{\mu \mu} -m_{\mu s}^{2} )} {m_{es}}
$
&
$\checkmark$
&
&
\\
(2,2,6)
&
$\frac{m_{\mu \tau}}{m_{\mu \mu}} = \frac{m_{\tau s}}{m_{\mu s}},
\frac{m_{\mu \mu}}{m_{\mu s}} = \frac{m_{\mu s}}{m_{ss}}
$
&
$\checkmark$
&
&
\\
(3,2,6)
&
$
\frac{m_{e \tau}}{m_{es}} =\frac{m_{\mu \tau}}{m_{\mu s}}= \frac{m_{\tau \tau}}{m_{\tau s}}=\frac{m_{\tau s}}{m_{ss}}=\sqrt{\frac{m_{\tau \tau}}{m_{ss}}}
$
&
$\checkmark$
&
$\checkmark$
&
\\
(4,2,6)
&
$\frac{m_{e \tau}}{m_{es}} = \frac{m_{\mu \tau}}{m_{\mu s}},
\frac{m_{\mu \mu}}{m_{\mu s}} = \frac{m_{\mu s}}{m_{ss}}
$
&
$\checkmark$
&
&
\\
(5,3,5)
&
$
\frac{m_{e \tau}}{m_{es}} =\frac{m_{\mu \tau}}{m_{\mu s}}= \frac{m_{\tau \tau}}{m_{\tau s}}=\frac{m_{\tau s}}{m_{ss}}=\sqrt{\frac{m_{\tau \tau}}{m_{ss}}}
$
&
$\checkmark$
&
$\checkmark
$
&
\\
\hline
Type& Scaling & $A_2$ & $\tilde{A}_2(\tilde M_R)$&$\tilde{A}_2(\tilde M_S)$ \\
\hline
(6,3,1)
&
$
\frac{(m_{\mu s}m_{\tau \tau} - m_{\mu \tau}m_{\tau s} )}{m_{e\mu}} = \frac{(m_{s s}m_{\tau \tau} -m_{\tau s}^{2} )} {m_{es}}
$
&
$\checkmark$
&
 &
\\
(8,2,6)
&
$\frac{m_{\mu \tau}}{m_{\tau \tau}} = \frac{m_{\mu s}}{m_{\tau s}},
\frac{m_{\tau \tau}}{m_{\tau s}}= \frac{m_{\tau s}}{m_{ss} }
$
&
$\checkmark$
&
&
\\
(7,3,5)
&
$
\frac{m_{e \mu}}{m_{es}} =\frac{m_{\mu \tau}}{m_{\tau s}}= \frac{m_{\mu \mu}}{m_{\mu s}}=\frac{m_{\mu s}}{m_{ss}}=\sqrt{\frac{m_{\mu \mu}}{m_{ss}}}
$
&
$\checkmark$
&
$\checkmark$
&
\\
(9,2,6)
&
$\frac{m_{e \mu}}{m_{es}} = \frac{m_{\mu \tau}}{m_{\tau s}},
\frac{m_{\tau \tau}}{m_{\tau s}}= \frac{m_{\tau s}}{m_{ss} }
$
&
$\checkmark$
&
 &
\\
(10,2,6)
&
$
\frac{m_{e \mu}}{m_{es}} =\frac{m_{\mu \tau}}{m_{\tau s}}= \frac{m_{\mu \mu}}{m_{\mu s}}=\frac{m_{\mu s}}{m_{ss}}=\sqrt{\frac{m_{\mu \mu}}{m_{ss}}}
$
&
$\checkmark$
&
$\checkmark$
&
\\
\hline
Type& Scaling &$B_3$ & $\tilde{B}_3(\tilde M_R)$&$\tilde{B}_3(\tilde M_S)$ \\
\hline
(11,2,6)
&
$
\frac{m_{e\tau}}{m_{es}} = \frac{m_{\mu\tau}}{m_{\mu s}} = \frac{m_{\tau\tau}}{m_{\tau s}} = \frac{m_{\tau s}}{m_{ss}} =\sqrt{\frac{m_{\tau \tau}}{m_{ss}}}
$
&
$\checkmark$
&
$\checkmark$
&
\\
(12,3,5)
&
$\frac{m_{e\tau}}{m_{es}} = \frac{m_{\mu\tau}}{m_{\mu s}} = \frac{m_{\tau\tau}}{m_{\tau s}} = \frac{m_{\tau s}}{m_{ss}} =\sqrt{\frac{m_{\tau \tau}}{m_{ss}}}
$
&
$\checkmark$
&
$\checkmark$
&
\\
(7,2,6)
&
$
\frac{m_{e \tau}}{m_{\tau s}} = \frac{m_{ee}}{m_{es}},
\frac{m_{ee}}{m_{es}} = \frac{ m_{es}}{m_{ss}}
$
&
$\checkmark$
&
 &
\\
(13,2,6)
&
$\frac{m_{e \tau}}{m_{es}} = \frac{m_{\mu \tau}}{m_{\mu s}},
\frac{m_{ee}}{m_{es}} = \frac{ m_{es}}{m_{ss}}
$
&
$\checkmark$
&
 &
\\
(14,3,1)
&
$
\frac{(m_{\tau \tau}m_{\mu s} - m_{\mu \tau}m_{\tau s} )}{m_{e\tau}} = \frac{(m_{e\tau}m_{\mu s} -m_{es}m_{\mu \tau} ) }{m_{ee}}
$
&
$\checkmark$
&
 &
\\
\hline
Type& Scaling&$B_4$ & $\tilde{B}_4(\tilde M_R)$&$\tilde{B}_4(\tilde M_S)$ \\
\hline
(15,2,6)
&
$
\frac{m_{e\mu}}{m_{es}} = \frac{m_{\mu\tau}}{m_{\tau s}} = \frac{m_{\mu\mu}}{m_{\mu s}} = \frac{m_{\mu s}}{m_{ss}} =\sqrt{\frac{m_{\mu \mu}}{m_{ss}}}
$
&
$\checkmark$
&
$\checkmark$
&
\\
(16,3,5)
&
$\frac{m_{e\mu}}{m_{es}} = \frac{m_{\mu\tau}}{m_{\tau s}} = \frac{m_{\mu\mu}}{m_{\mu s}} = \frac{m_{\mu s}}{m_{ss}} =\sqrt{\frac{m_{\mu \mu}}{m_{ss}}}
$
&
$\checkmark$
&
$\checkmark$
&
\\
(5,2,6)
&
$
\frac{m_{e \mu}}{m_{\mu s}} = \frac{m_{ee}}{m_{es}},
\frac{m_{ee}}{m_{es}} = \frac{m_{es}}{m_{ss}}
$
&
$\checkmark$
&
 &
\\
(17,2,6)
&
$\frac{m_{e \mu}}{m_{es}} = \frac{m_{\mu \tau}}{m_{\tau s}},
\frac{m_{ee}}{m_{es}} = \frac{ m_{es}}{m_{ss}}
$
&
$\checkmark$
&
 &
\\
(14,3,1)
&
$
\frac{(m_{\mu \mu}m_{\tau s} - m_{\mu \tau}m_{\mu s} )}{m_{e\mu}} = \frac{(m_{e\mu}m_{\tau s} -m_{es}m_{\mu \tau} ) }{m_{ee}}
$
&
$\checkmark$
&
 &
\\
\hline
Type& Scaling&$D_1$ & $\tilde{D}_1(\tilde M_R)$&$\tilde{D}_1(\tilde M_S)$
 \\
\hline
(19,3,1)
&
$
\frac{m_{ee}}{m_{e\tau}} - \frac{m_{e\tau}}{m_{\tau \tau}} = \left( \frac{m_{ee}}{m_{\tau \tau}} - \frac{m_{e\tau}^2}{m_{\tau \tau}^2} \right) \frac{\ell}{c}
$
&
$\checkmark$
&

&
$\checkmark$
\\
(20,2,6)
&
$\frac{m_{e \mu}}{m_{\mu s}} = \frac{m_{e\tau}}{m_{\tau \tau}},
\frac{m_{\tau \tau}}{m_{\tau s}} = \frac{m_{\tau s}}{m_{ss}}
$
&
$\checkmark$
&
 &
\\
(2,3,5)
&
$
\frac{m_{e \mu}}{m_{\mu s}} = \frac{m_{e\tau}}{m_{\tau s}}=\sqrt{\frac{m_{ee}}{m_{ss}}}
$
&
$\checkmark$
&
$\checkmark$
&
\\
(21,2,6)
&
$\frac{m_{e e}}{m_{es}} = \frac{m_{es}}{m_{ss}}=\frac{m_{e\mu}}{m_{\mu s}}=\frac{m_{e\tau}}{m_{\tau s}}=\sqrt{\frac{m_{ee}}{m_{ss}}}
$
&
$\checkmark$
&
$\checkmark$
&
\\
(16,2,6)
&
$
\frac{m_{e \tau}}{m_{\tau s}} = \frac{m_{es}}{m_{ss}},
\frac{m_{ee}}{m_{es}} = \frac{m_{e s}}{m_{ss}}
$
&
$\checkmark$
&
 &
\\
\hline
\end{tabular}
\label{table_scaling}}
\end{table}
\begin{table}[t]
\tbl{Scaling relations in the MES. (Cont.) }
{\begin{tabular}{|c|c|c|c|c|}
\hline
Type& Scaling & $D_2$ & $\tilde{D}_2(\tilde M_R)$ & $\tilde{D}_2(\tilde M_S)$ \\
\hline
(23,3,1)
&
$
\frac{m_{ee}}{m_{e\mu}} = \frac{m_{e\mu}}{m_{\mu \mu}} = \left( \frac{m_{ee}}{m_{\mu \mu}} - \frac{m_{e\mu}^2}{m_{\mu \mu}^2} \right) \frac{f}{c}
$
&
$\checkmark$
&
$$
&
$\checkmark$
\\
(22,9,6)
&
$\frac{m_{e \tau}}{m_{\tau s}} = \frac{m_{e\mu}}{m_{\mu \mu}},
\frac{m_{\mu \mu}}{m_{\mu s}} = \frac{m_{\mu s}}{m_{ss}}
$
&
$\checkmark$
&
&
\\
(8,3,5)
&
$
\frac{m_{e \mu}}{m_{\mu s}} = \frac{m_{e\tau}}{m_{\tau s}}=\sqrt{\frac{m_{ee}}{m_{ss}}}
$
&
$\checkmark$
&
$\checkmark$
&
\\
(24,2,6)
&
$\frac{m_{e e}}{m_{es}} = \frac{m_{es}}{m_{ss}}=\frac{m_{e\mu}}{m_{\mu s}}=\frac{m_{e\tau}}{m_{\tau s}}=\sqrt{\frac{m_{ee}}{m_{ss}}}
$
&
$\checkmark$
&
$\checkmark$
&
\\
(16,2,6)
&
$
\frac{m_{e \mu}}{m_{\mu s}} = \frac{m_{es}}{m_{ss}},
\frac{m_{ee}}{m_{\tau s}} = \frac{m_{\tau s}}{m_{ss}}
$
&
$\checkmark$
&
&
\\
\hline
Type& Scaling&$F_1$ & $\tilde{F}_1(\tilde M_R)$&$\tilde{F}_1(\tilde M_S)$ \\
\hline
(24,3,2)
&
$
\frac{m_{ee}}{m_{es}}=\frac{m_{es}}{m_{ss}}
$
&
$\checkmark$
&
&
\\
(21,3,2)
&
$
\frac{m_{ee}}{m_{es}}=\frac{m_{es}}{m_{ss}}
$
&
$\checkmark$
&
&
\\
(17,1,3)
&
$
\frac{m_{e \tau}}{m_{\tau \tau}} = \frac{m_{\mu s}}{m_{\tau s}}
$
&
$\checkmark$
&
&
\\
(21,2,6)
&
$
\frac{m_{\mu \mu}}{m_{\mu \tau}} = \frac{m_{\mu s}}{m_{\tau s}}
$
&
$\checkmark$
&
&
\\
\hline
Type& Scaling&$F_2$ & $\tilde{F}_2(\tilde M_R)$&$\tilde{F}_2(\tilde M_S)$ \\
\hline
(24,3,2)
&
$
\frac{m_{\mu \mu}}{m_{\mu s}}=\frac{m_{\mu s}}{m_{\mu \mu}}
$
&
$\checkmark$
&
&
\\
(21,3,2)
&
$
\frac{m_{\mu \mu}}{m_{\mu s}}=\frac{m_{\mu s}}{m_{\mu \mu}}
$
&
$\checkmark$
&
&
\\
(17,1,3)
&
$
\frac{m_{e \tau}}{m_{\tau \tau}} = \frac{m_{\mu s}}{m_{\tau s}}
$
&
$\checkmark$
&
&
\\
(21,2,6)
&
$
\frac{m_{\mu \mu}}{m_{\mu \tau}} = \frac{m_{\mu s}}{m_{\tau s}}
$
&
$\checkmark$
&
&
\\
\hline
Type& Scaling&$F_3$ & $\tilde{F}_3(\tilde M_R)$&$\tilde{F}_3(\tilde M_S)$ \\
\hline
(11,3,2)
&
$
\frac{m_{\tau \tau}}{m_{\tau s}}=\frac{m_{\tau s}}{m_{ss}}
$
&
$\checkmark$
&
&
\\
(10,3,2)
&
$
\frac{m_{\tau \tau}}{m_{\tau s}}=\frac{m_{\tau s}}{m_{ss}}
$
&
$\checkmark$
&
&
\\
(25,1,2)
&
$
\frac{m_{ee}}{m_{es}} = \frac{m_{e\mu}}{m_{\mu s}}
$
&
$\checkmark$
&
&
\\
(26,1,2)
&
$
\frac{m_{e \mu}}{m_{es}} = \frac{m_{\mu \mu}}{m_{\mu s}}
$
&
$\checkmark$
&
&
\\
\hline
\end{tabular}
\label{scalingcon}}
\end{table}

We now understand that vanishing $m_{ee}$ and $m_{e\mu}$ are not necessary to maintain the scaling relation $\left(M^{(3)}_D, \tilde M^{(2)}_R, M^{(6)}_S\right)$ in the $\tilde A_1$ texture. This raises the question: how large $|m_{ee}|$ and $|m_{e\mu}|$ are allowed with the scaling relation $\left(M^{(3)}_D, \tilde M^{(2)}_R, M^{(6)}_S\right)$? To answer this question, we performed a supplemental numerical calculation. 

To estimate the magnitude of the matrix element using the neutrino data, we assume the charged lepton mass matrix to be diagonal and real. Under this assumption, the $4 \times 4$ Majorana neutrino mass matrix is obtained as follows:
\begin{align}
M_\nu=UM_{diag}U^T,
\end{align}
where $U$ denotes $4\times 4$ lepton mixing matrix, and $M_{diag}= {\rm diag} (m_1, m_2, m_3, m_4)$ denotes the diagonal neutrino mass matrix. The lepton mixing matrix $U$ contains the neutrino mixing angles $\theta_{12}, \theta_{13}, \theta_{14}, \theta_{23}, \theta_{24}, \theta_{34}$, the Dirac phases $\delta_{13}, \delta_{14}, \delta_{24}$ and the Majorana phases $\alpha, \beta, \gamma$. We parametrize $U$ as  \cite{Goswami2006PRD}
\begin{align}
U=R_{34}\Tilde{R}_{24}\Tilde{R}_{14}R_{23}\Tilde{R}_{13}R_{12} P
=
\begin{pmatrix}
U_{e1}&U_{e2}&U_{e3}&U_{e4} \\
U_{\mu 1}&U_{\mu 2}&U_{\mu 3}&U_{\mu 4} \\
U_{\tau 1}&U_{\tau 2}&U_{\tau 3}&U_{\tau 4} \\
U_{s1}&U_{s2}&U_{s3}&U_{s4} \\
\end{pmatrix},
\end{align}
where $R_{ij}$ denotes the rotation matrix in the $ij$ generation space, i.e., 
\begin{align}
    R_{34} =
    \begin{pmatrix}
        1&0&0&0 \\
        0&1&0&0 \\
        0&0&c_{34}&s_{34} \\
        0&0&-s_{34}&c_{34}
    \end{pmatrix},
    \quad
     \textcolor{black}{\Tilde{R}_{24}=
    \begin{pmatrix}
        1&0&0&0 \\
        0&c_{24}&0&s_{24}e^{-i\delta_{24}} \\
        0&0&1&0 \\
        0&-s_{24}e^{i\delta_{24}}&0&c_{24}
    \end{pmatrix}},
\end{align}
with the usual notation $s_{ij} = \sin\theta_{ij}$ and $c_{ij} = \cos\theta_{ij}$. The diagonal matrix $P$ is defined as
\begin{align}
    P = {\rm diag}\left( 1,e^{-i\alpha/2},e^{-i(\beta/2-\delta_{13})},e^{-i(\gamma /2-\delta_{14})} \right).
\end{align}

The current neutrino data \cite{Nufit53 ,Gariazzo2016 ,Giunti2017, Kopp2011} show the following best-fit value:
\begin{align}\label{neutrinopara}
& \Delta m_{21}^2 [ 10^{-5} {\rm eV}^2] =  7.49 \quad (6.92 - 8.05), \nonumber \\
& \Delta m_{31}^2 [ 10^{-3} {\rm eV}^2] ({\rm NO})=  2.513 \quad (2.451 - 2.578), \nonumber \\
& -\Delta m_{32}^2 [ 10^{-3} {\rm eV}^2] ({\rm IO})=  2.484 \quad (2.421 - 2.547), \nonumber\\
&\Delta m_{41}^2[{\rm eV}^2] ({\rm NO})=  1.63 \quad (0.87 - 2.04),  \nonumber\\
&\Delta m_{43}^2[{\rm eV}^2] ({\rm IO})= 1.63 \quad (0.87 - 2.04),  \nonumber\\
&\sin^2\theta_{12} = 0.308  \quad (0.275 - 0.344), \nonumber\\
&\sin^2\theta_{23}({\rm NO}) = 0.470\quad (0.435-0.585),  \nonumber\\
&\sin^2\theta_{23} ({\rm IO}) = 0.550 \quad (0.440 - 0.584), \nonumber\\
&\sin^2\theta_{13} ({\rm NO}) = 0.02215 \quad (0.02030-0.02388), \nonumber\\
&\sin^2\theta_{13} ({\rm IO}) = 0.02231  \quad(0.02060 - 0.02409), \nonumber\\
&|U_{e4}|^2 = 0.027 \quad (0.012-0.047),\nonumber \\
&|U_{\mu 4}|^2 = 0.013 \quad (0.005-0.03),\nonumber \\
&|U_{\tau 4}|^2 < 0.16,
\end{align}
where NO indicates the normal mass ordering $(m_1 < m_2 < m_3)$, and IO indicates the inverted mass ordering ($m_3 < m_1 \lesssim m_2$) of the active neutrinos. The parentheses denote the $3 \sigma$ region. We set $m_1 = 0$ for NO and $m_3=0$ for IO. According to Kumar and Patgiri\cite{Kumar2020NPB}, the range of the Dirac and Majorana phases are considered within $(0 - 2\pi)$, and $\sin\theta_{34}$ is taken within $(0-0.4)$. We randomly make $10^8$ parameter sets of $\{m_1,m_2,m_3,m_4,s_{12},s_{23},s_{13},s_{14}$,$s_{24},s_{34},\delta_{13},\delta_{14},\delta_{24},\alpha,\beta,\gamma\}$ with 3 $\sigma$ values in Eq. (\ref{neutrinopara}) to our numerical calculations.

\begin{figure}[t]
\centering
\begin{minipage}[b]{0.49\columnwidth}
    \centering
    \includegraphics[width=1.4\columnwidth]{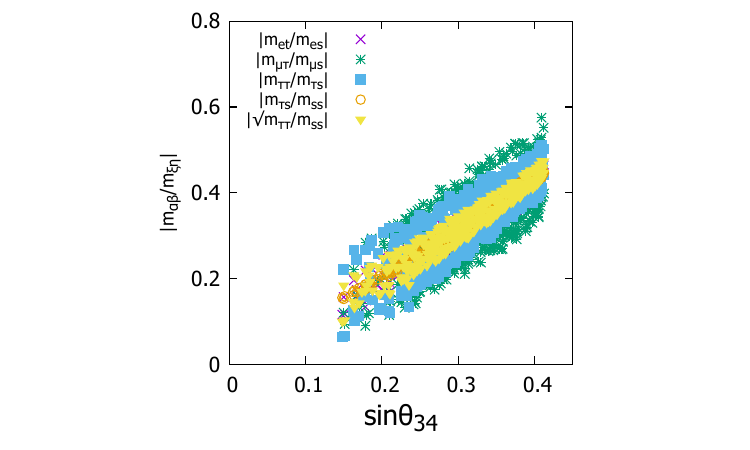}
\end{minipage}
\begin{minipage}[b]{0.49\columnwidth}
    \centering
    \includegraphics[width=1.4\columnwidth]{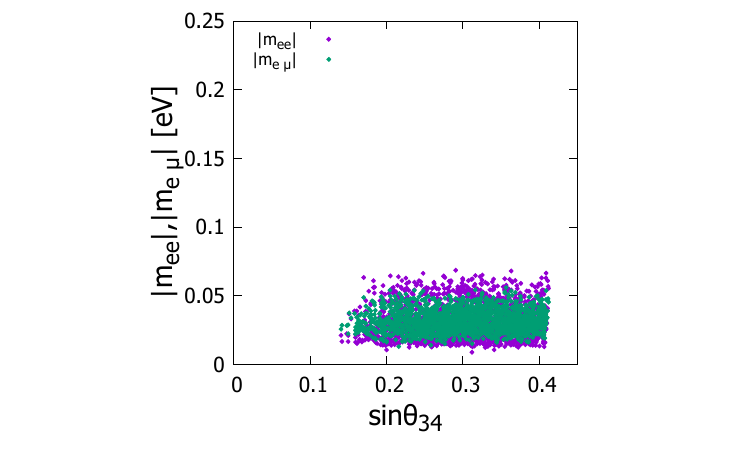}
\end{minipage}
\caption{Left: ratios of the two matrix elements in $\tilde{A}_1$ texture. Right: $|m_{ee}|$ and $|m_{e\mu}|$, which are picked up from the overlap region in the left panel.}
\label{A1scaling}
\end{figure}

\begin{table}[t]
\tbl{Upper limit of the elements that \textcolor{black}{were originally vanishing} in the \textcolor{black}{two-zero texture} schemes in Eq.(\ref{Eq:textureTwoZero}).\label{scalingmab} }
{\begin{tabular}{|c|c|c|c|}
\hline
 &Textures&$|\textcolor{black}{m}_{\alpha\alpha}|$[eV] &$|{\color{black}m}_{\alpha\beta}|$[eV]\\
 \hline
 &$\tilde{A}_1$&$ 0.0705 $ &$ 0.0592$\\
 &$\tilde{A}_2$&$ 0.0700 $ &$ 0.1253 $\\
 &$\tilde{B}_3$&$ 0.0723 $ &$ 0.0575 $\\
 NO&$\tilde{B}_4$&$ 0.2495 $ &$ 0.1222$\\
 &$\tilde{D}_1(2,3,5)$&$ 0.0728$ &$ 0.1256$\\
 &$\tilde{D}_1(21,2,6)$&$ 0.0728$ &$ 0.1256$\\
 &$\tilde{D}_2(8,3,5)$&$ 0.2548$ &$ 0.1256$\\
 &$\tilde{D}_2(24,2,6)$&$ 0.2548$ &$ 0.1256$\\
\hline
& \textcolor{black}{T}extures & $|{\color{black}m}_{\alpha\alpha}|$[eV] & $|{\color{black}m}_{\alpha\beta}|$[eV] \\
\hline
 &$\tilde{A}_1$&$ 0.1131$ &$ 0.0797$\\
 &$\tilde{A}_2$&$ 0.1128$ &$ 0.1455$\\
 &$\tilde{B}_3$&$ 0.0702$ &$ 0.0797$\\
IO &$\tilde{B}_4$&$ 0.2525$ &$ 0.1400$\\
 &$\tilde{D}_1(2,3,5)$&$ 0.0702$ &$ 0.1200$\\
 &$\tilde{D}_1(21,2,6)$&$ 0.1031$ &$ 0.0700$\\
 &$\tilde{D}_2(8,3,5)$&$ 0.2510$ &$ 0.1200$\\
 &$\tilde{D}_2(24,2,6)$&$ 0.2302$ &$ 0.1031$\\
\hline
\end{tabular}
}
\end{table}

\begin{table}[t]
 \tbl{\textcolor{black}{Constraints on CP phases and  $\sin \theta_{34}$ for NO in the two-zero texture schemes in Ref.\cite{Kumar2020NPB}}}
{\begin{tabular}{|c|c|}
\hline
 Textures&$ \delta_{13},  \delta_{14}, \delta_{24}$\\
 \hline
 $A_2$ Type(7,3,5),(10,2,6) &$ \delta_{13}=(45^{\circ}-90^{\circ})$ ,$\delta_{14}= (180^{\circ}-225^{\circ}), \delta_{24}=(180^{\circ}-225^{\circ}) $  \\
 $D_1$  Type(2,3,5)&$ \delta_{13}=(135^{\circ}-180^{\circ}), \delta_{24}=(270^{\circ}-300^{\circ}),\delta_{14}= (0^{\circ}-30^{\circ})$ \\
 $D_1$ Type(21,2,6)&$ \delta_{13}=(45^{\circ}-90^{\circ}), \delta_{24}=\delta_{14}= (270^{\circ}-315^{\circ})$ \\
\hline
 Textures & $\alpha, \beta , \gamma$ \\
\hline
 $A_2$ Type(7,3,5),(10,2,6) & $\alpha=(135^{\circ}-90^{\circ}),
\beta=(0^{\circ}-30^{\circ}), 
\gamma=(315^{\circ}-360^{\circ}) $\\
$D_1$  Type(2,3,5)&$\alpha=(180^{\circ}-225^{\circ}),
\beta=(90^{\circ}-135^{\circ})$ \\
 $D_1$ Type(21,2,6)&$\alpha=(135^{\circ}-180^{\circ}),
\beta=(0^{\circ}-30^{\circ}), 
\gamma=(225^{\circ}-315^{\circ})$ \\
\hline
Textures & $\sin \theta_{34}$ \\
\hline
 $A_2$ Type(7,3,5),(10,2,6) & $\sin \theta_{34}  > 0.04$ \\
$D_1$  Type(2,3,5) & $\sin \theta_{34}  > 0$\\
$D_1$ Type(21,2,6) & $\sin \theta_{34}  > 0 $\\
\hline
\end{tabular}
\label{twozeroCPsin34}}
\end{table}

Let us estimate the allowed magnitude of $|m_{ee}|$ and $|m_{e\mu}|$ with the scaling relations $\left(M^{(3)}_D, \tilde M^{(2)}_R, M^{(6)}_S\right)$ in $\tilde{A}_1$ texture. First, we plot the ratios of the two matrix elements in $\tilde{A}_1$ texture for NO, as shown in the left panel of Fig. \ref{A1scaling}. The overlapping points in the panel are regarded within the scaling relation $\left(M^{(3)}_D, \tilde M^{(2)}_R, M^{(6)}_S\right)$ in $\tilde{A}_1$ texture. In this study, we require that the overlapped points satisfy the criteria of ${\rm Re}(m_{ab}/m_{cd}-m_{ef}/m_{gh}) \le 0.07$ and ${\rm Im}(m_{ab}/m_{cd}-m_{ef}/m_{gh}) \le 0.07$. We choose ``0.07" as a realistic number for the numerical calculation by the laptop PC that we used. Then, we picked up the overlapped $|m_{ee}|$ and $|m_{e\mu}|$, as shown in the right panel in Fig. \ref{A1scaling}. The upper limits of $|m_{ee}|$ and $|m_{e\mu}|$ within scaling relation $\left(M^{(3)}_D, \tilde M^{(2)}_R, M^{(6)}_S\right)$ in $\tilde{A}_1$ texture for NO are shown on the top row in Table \ref{scalingmab}, e.g., $|m_{ee}| \le 0.0705$ eV and $|m_{e\mu}| \le 0.0592$ eV.  Table \ref{scalingmab} also shows the upper limit of the matrix elements in other textures that were originally \textcolor{black}{vanishing} in the texture two zero scheme in Eq.(\ref{Eq:textureTwoZero}).

{\color{black}To make this study more complete, we present some comparisons between the previous studies and this study. 

First, we compare the previous studies in Refs \cite{Nath2017JHEP,Patgiri2017IJMPA} and this study. In the previous study, the IO is only allowed pattern of the neutrino mass hierarchy in their zero texture framework. On the contrary, both hierarchies, NO and IO, are possible for the generalized neutrino mass matrices in this study. 

Second, we show the comparison between the studies by Kumar and Patgiri\cite{Kumar2020NPB}, and by us. Table \ref{twozeroCPsin34} shows that the constraints on CP phases and  $\sin \theta_{34}$ for NO in the zero texture scheme in Ref.\cite{Kumar2020NPB}. On the other hand, we show the constraints on CP phases and $\sin \theta_{34}$ for NO in the generalized neutrino mass matrix scheme in this study in Table \ref{tildeCPsin34}. Comparing the two tables, we see that the wider ranges of CP phases are allowed in the generalized mass matrices. This observation is remarkable for the phases $\alpha$, $\beta$ and $\gamma$. For an active-sterile mixing angle $\theta_{34}$, the prediction of it from the generalized mass matrices is somewhat more restrictive than that from the texture zero mass matrices.}

{\color{black}\begin{table}[t]
 \tbl{\color{black}Constraints on CP phases and  $\sin \theta_{34}$ for NO in the generalized neutrino mass matrix schemes in this study.}
{\begin{tabular}{|c|c|}
\hline
 Textures&$ \delta_{13},  \delta_{14}, \delta_{24}$\\
 \hline
 $\tilde A_2$ Type(7,3,5),(10,2,6) &$ \delta_{13}=(0^{\circ}-360^{\circ}),\delta_{14}= (0^{\circ}-360^{\circ}), \delta_{24}=(0^{\circ}-125^{\circ},219^{\circ}-360^{\circ}) $  \\
 $\tilde D_1$  Type(2,3,5)&$ \delta_{13}=(0^{\circ}-360^{\circ}),\delta_{14}= (0^{\circ}-360^{\circ}),\delta_{24}=(0^{\circ}-360^{\circ})$ \\
 $\tilde D_1$ Type(21,2,6)&$ \delta_{13}=(0^{\circ}-360^{\circ}),\delta_{14}= (0^{\circ}-360^{\circ}),\delta_{24}=(0^{\circ}-90^{\circ},270^{\circ}-360^{\circ})$ \\
\hline
 Textures & $\alpha, \beta , \gamma$ \\
\hline
 $\tilde A_2$ Type(7,3,5),(10,2,6) & $\alpha=(0^{\circ}-360^{\circ}),
\beta=(0^{\circ}-360^{\circ}), 
\gamma=(0^{\circ}-360^{\circ}) $\\
$\tilde D_1$  Type(2,3,5)&$\alpha=(0^{\circ}-360^{\circ}),
\beta=(0^{\circ}-360^{\circ}),\gamma=(0^{\circ}-360^{\circ})$ \\
 $\tilde D_1$ Type(21,2,6)&$\alpha=(0^{\circ}-360^{\circ}),
\beta=(0^{\circ}-360^{\circ}), 
\gamma=(0^{\circ}-360^{\circ})$ \\
\hline
Textures & $\sin \theta_{34}$ \\
\hline
 $\tilde A_2$ Type(7,3,5),(10,2,6) & $\sin \theta_{34}  > 0.126$ \\
$\tilde D_1$  Type(2,3,5) & $\sin \theta_{34}  > 0.006$\\
$\tilde D_1$ Type(21,2,6) & $\sin \theta_{34}  > 0.008 $\\
\hline
\end{tabular}
\label{tildeCPsin34}}
\end{table}}
\section{Summary\label{section:summary}}
The scaling relations in the two-zero textures of the neutrino mass matrices in the MES mechanism were discovered by Kumar and Patgiri. In this study, we found that some scaling relations in these two-zero textures can be satisfied without requiring two zero elements in the texture. A scaling that is independent of the two-zero texture can be considered more general than restrictive scaling, which requires two zero elements in the mass matrix. Thus, our study increased the value of the discovery by Kumar and Patgiri.




\begin{thebibliography}{0}
\bibitem{Frizsch2020PPNP}
H. Fritzsch and Z. Xing, \Journal{\PPNP}{45}{1}{2020}.
%
\bibitem{Altarelli2010RMP}
G. Altarelli and F. Feruglio, \Journal{\RMP}{82}{2701}{2010}.
%
\bibitem{Ishimori2010PTPS}
H. Ishimori, T. Kobayashi, H. Ohki, Y. Shimizu, H. Okada, and M. Tanimoto, \Journal{\PTPS}{183}{1}{2010}
%
\bibitem{Xing2020PREP}
Z. Xing, \Journal{\PREP}{854}{1}{2020}.
%
\bibitem{Xing2023RPP}
Z. Xing, \Journal{\RPP}{86}{076201}{2023}.
%
\bibitem{Mohapatra2007PLB}
R. N. Mohapatra and W. Rodejohann, \Journal{\PLB}{644}{59}{2007}.
%
\bibitem{Blum2007PRD}
A. Blum, R. N. Mohapatra, and W. Rodejohann, \Journal{\PRD}{76}{053003}{2007}.
%
\bibitem{Joshipura2009PLB}
A. S. Joshipura and W. Rodejohann, \Journal{\PLB}{678}{276}{2009}.
%
\bibitem{Yasur2012PRD}
M. Yasu{\`e}, \Journal{\PRD}{86}{116011}{2012}.
%
\bibitem{Samanta2016EPJC}
R. Samanta, P. Roy, and A. Ghosal, \Journal{\EPJC}{76}{662}{2016}.
%
\bibitem{Sinha2017JHEP}
R. Sinha, R. Samanta, and A. Ghosal, \Journal{\JHEP}{12}{030}{2017}.
%
\bibitem{Barry2011JHEP}
J. Barry, W. Rodejohann, and H. Zhang, \Journal{\JHEP}{1107}{091}{2011}.
%
\bibitem{Kumar2020NPB}
P. Kumar and M. Patgiri, \Journal{\NPB}{957}{115082}{2020}.
%
\bibitem{Nath2017JHEP}
N. Nath, M. Ghosh, S. Goswamia, and S. Gupta \Journal{\JHEP}{03}{075}{2017}.
%
\bibitem{Patgiri2017IJMPA}
M. Patgiri, P. Kumar, and D. Sarma, \Journal{\IJMPA}{32}{27}{2017}{1750168}.
%
\bibitem{Sarma2019EPJC}
N. Sarma, K. Bora, and D. Borah, \Journal{\EPJC}{79}{129}{2019}.
%
\bibitem{Das2019NPB}
P. Das, A. Mukherjee, and M. K. Das, \Journal{\NPB}{941}{755-779}{2019}.
%
\bibitem{Patgiri2019IJMPA}
M. Patgiri and P. Kumar, \Journal{\IJMPA}{34}{11}{2019}{1950059}.
%
\bibitem{Das2022NPB}
P. Das, M. K. Das, and N. Khan, \Journal{\NPB}{980}{115810}{2022}.
%
\bibitem{Goswami2006PRD}
S. Goswami and W. Rodejohann, \Journal{\PRD}{73}{113003}{2006}.
%

\bibitem{Nufit53}
I. Esteban, M.C. Gonzalez-Garcia, M. Maltoni, I. Martinez-Soler, J. Paulo Pinheiro, and T. Schwetz,  arXiv:2410.05380. http://www.nu-fit.org
%
\bibitem{Gariazzo2016}
S Gariazzo, C Giunti, M Laveder, Y F Li and E M Zavaninl, \Journal{\JPG}{43}{033001}{2016}.
%

\bibitem{Giunti2017}
C. Giunti, \Journal{\JPCS}{888}{012231}{2017}.
%
\bibitem{Kopp2011}
J. Kopp, M. Maltoni, and T. Schwetz, \Journal{\PRL}{107}{091801}{2011}.

\end{thebibliography}
\end{document}